\documentclass{article}
\usepackage{smc2019}
\usepackage{times}
\usepackage{ifpdf}
\usepackage[english]{babel}
\usepackage{cite}
\usepackage{amsmath} % popular packages from Am. Math. Soc. Please use the 
\usepackage{amssymb} % related math environments (split, subequation, cases,
\usepackage{amsfonts}% multline, etc.)
\usepackage{bm}      % Bold Math package, defines the command \bf{}
\usepackage[font=footnotesize]{subfig}

\usepackage{multicol}

\def\papertitle{Autoencoders for music sound modeling: a comparison of linear, shallow, deep, recurrent and variational models}
\def\firstauthor{Roche}

\newif\ifpdf
\ifx\pdfoutput\relax
\else
   \ifcase\pdfoutput
      \pdffalse
   \else
      \pdftrue
\fi

\ifpdf % compiling with pdflatex
	\usepackage{hyperref}
	
  \usepackage[pdftex]{graphicx}
  \DeclareGraphicsExtensions{.pdf,.jpeg,.png}

  \usepackage[figure,table]{hypcap}

\else % compiling with latex
  \usepackage[dvips,
    bookmarksnumbered, % use section numbers with bookmarks
    pdfstartview=XYZ % start with zoom=100% instead of full screen
  ]{hyperref}  % hyperrefs are active in the pdf file after conversion

  \usepackage[dvips]{epsfig,graphicx}
  \graphicspath{{./figures/}}
  \DeclareGraphicsExtensions{.eps}

  \usepackage[figure,table]{hypcap}
\fi

\hypersetup{
    colorlinks,%
    citecolor=black,%
    filecolor=black,%
    linkcolor=black,%
    urlcolor=black
}

\title{\papertitle}

\oneauthor
   {Fanny Roche\textsuperscript{\normalfont 1,3} \qquad Thomas Hueber\textsuperscript{\normalfont 3} \qquad Samuel Limier\textsuperscript{\normalfont 1} \qquad Laurent Girin\textsuperscript{\normalfont 2,3} } {\textsuperscript{1}Arturia, Meylan, France \qquad \textsuperscript{2}Inria Grenoble Rh\^one-Alpes, France \\ \textsuperscript{3}Univ. Grenoble Alpes, CNRS, Grenoble INP, GIPSA-lab, Grenoble, France \\ %
     {\tt \href{mailto:fanny.roche@gipsa-lab.fr}{fanny.roche@gipsa-lab.fr}}}

\begin{document}
\capstartfalse
\maketitle
\capstarttrue
\begin{abstract}
This study investigates the use of non-linear unsupervised dimensionality reduction techniques to compress a music dataset into a low-dimensional representation which can be used in turn for the synthesis of new sounds. We systematically compare (shallow) autoencoders (AEs), deep autoencoders (DAEs), recurrent autoencoders (with Long Short-Term Memory cells -- LSTM-AEs) and variational autoencoders (VAEs) with principal component analysis (PCA) for representing the high-resolution short-term magnitude spectrum of a large and dense dataset of music notes into a lower-dimensional vector (and then convert it back to a magnitude spectrum used for sound resynthesis). Our experiments were conducted on the publicly available multi-instrument and multi-pitch database NSynth. Interestingly and contrary to the recent literature on image processing, we can show that PCA systematically outperforms shallow AE. Only deep and recurrent architectures (DAEs and LSTM-AEs) lead to a lower reconstruction error. The optimization criterion in VAEs being the sum of the reconstruction error and a regularization term, it naturally leads to a lower reconstruction accuracy than DAEs but we show that VAEs are still able to outperform PCA while providing a low-dimensional latent space with nice ``usability'' properties. We also provide corresponding objective measures of perceptual audio quality (PEMO-Q scores), which generally correlate well with the reconstruction error.
\end{abstract}

\section{Introduction}\label{sec:introduction}

% 0. What is music synthesis?
%Many different methods for sound texture generation in music exist and are commonly used in synthesizers 
%For a few years, researchers have started to investigate new synthesis methods using data-driven machine learning techniques, in particular artificial neural networks (ANNs) \cite{sarroff2014, CooperUnion, NSynth}.

%---------
Deep neural networks, and in particular those trained in an unsupervised (or self-supervised) way such as autoencoders \cite{Hinton2006} or GANs  \cite{Goodfellow2014}, have shown nice properties to extract latent representations from large and complex datasets. Such latent representations can be sampled to generate new data. These types of models are currently widely used for image and video generation \cite{reed2016, Mehri2017, tulyakov2018mocogan}. In the context of a project aiming at designing a music sound synthesizer driven by high-level control parameters and propelled by data-driven machine learning, we investigate the use of such techniques for music sound generation as an alternative to classical music sound synthesis techniques like additive synthesis, subtractive synthesis, frequency modulation, wavetable synthesis or physical modeling \cite{MirandaBook}. 

So far, only a few studies in audio processing have been proposed in this line, with a general principle that is similar to image synthesis/transformation: projection of the signal space into a low-dimensional latent space (encoding or embedding), modification of the latent coefficients, and inverse transformation of the modified latent coefficients into the original signal space (decoding). 

In \cite{sarroff2014,CooperUnion}, the authors implemented this principle with autoencoders to process normalized magnitude spectra. An autoencoder (AE) is a specific type of artificial neural network (ANN) architecture which is trained to reconstruct the input at the output layer, after passing through the latent space. Evaluation was made by computing the mean squared error (MSE) between the original and the reconstructed magnitude spectra. 
In \cite{NSynth}, NSynth, an audio synthesis method based on a time-domain autoencoder inspired from the WaveNet speech synthesizer \cite{Oord_wavenet2016} was proposed. The authors investigated the use of this model to find a high-level latent space well-suited for interpolation between instruments. Their autoencoder is conditioned on pitch and is fed with raw audio from their large-scale multi-instrument and multi-pitch database (the NSynth dataset). 
%Here again, the evaluation was only qualitative and the NSynth model was not compared to classical autoencoders or PCA. Also, 
% This model may be difficult to train for non-experts and requires a lot of computational power.
This approach led to promising results but has a high computational cost.

Another technique to synthesize data using deep learning is the so-called variational autoencoder (VAE) originally proposed in \cite{Kingma2013}, which is now popular for image generation. A VAE can be seen as a probabilistic/generative version of an AE. Importantly, in a VAE, a prior can be placed on the distribution of the latent variables, so that they are well suited for the control of the generation of new data.
%Also in a VAE, a target distribution of the latent variables can be set so that ideally each latent variable may control more clearly one single perceptual signal dimension. All this makes VAEs well suited for signal synthesis and they 
This has been recently exploited for the modeling and transformation of speech signals \cite{blaauw2016modeling, hsu2017learning} and also for music sounds synthesis \cite{Esling2018}, incorporating some fitting of the latent space with a perceptual timbre space. VAEs have also been recently used for speech enhancement \cite{leglaive2018variance,leglaive2019semi,li2019fast}.
%Recent studies in speech \cite{blaauw2016modeling, hsu2017learning} and music \cite{Esling2018} suggest that VAE can be adapted to extract latent representation that correlate with some perceptual dimensions (e.g. in gender and phoneme identity). 

\begin{figure*}[!htb]
\centering
\begin{minipage}{0.7\textwidth}
	\includegraphics[width=\textwidth]{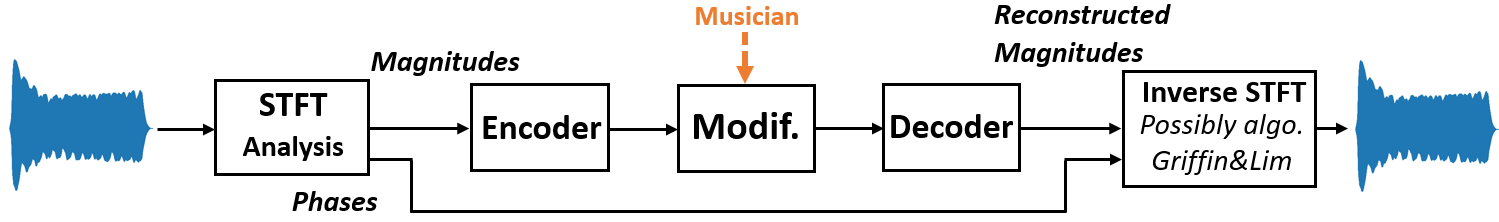}
	\caption{Global diagram of the sound analysis-transformation-synthesis process.}
	\label{fig:schema}
\end{minipage}
\end{figure*}

In line with the above-presented studies, the goal of the present paper is i) to provide an extensive comparison of several autoencoder architectures including shallow, deep, recurrent and variational autoencoders, with a systematic comparison to a linear dimensionality reduction technique, in the present case Principal Component Analysis (PCA) (to the best of our knowledge, such comparison of non-linear approaches with a linear one has never been done in previous studies). This is done using both an objective physical measure (root mean squared error -- RMSE) and an objective perceptual measure (PEMO-Q \cite{huber2006pemo}); ii) to compare the properties of the latent space in terms of correlation between the extracted dimensions; and iii) to illustrate how interpolation in the latent space can be performed to create interesting hybrid sounds.

\section{Methodology}\label{sec:method}

The global methodology applied 
%to extract a compact high-level space from a database of multi-instrument and multi-pitch music sounds 
for (V)AE-based analysis-transformation-synthesis of audio signals in this study is in line with previous works \cite{sarroff2014,CooperUnion,blaauw2016modeling, hsu2017learning}. It is illustrated in Fig.~\ref{fig:schema} and is described in the next subsections.

\subsection{Analysis-Synthesis}\label{sec:analysis}

First, a Short-Term Fourier Transform (STFT) analysis is performed on the input audio signal. 
% note pour Fanny: STFT includes the sliding-window processing. Therefore ``sliding-window STFT'' is redundant. Fourier analysis without sliding window is simply DFT analysis or FFT analysis.
The magnitude spectra are sent to the model (encoder input) on a frame-by-frame basis, and the phase spectra are stored for the synthesis stage.
After possible modifications of the extracted latent variables (at the bottleneck layer output, see next subsection), the output magnitude spectra is provided by the decoder. The output audio signal is synthesized by combining the decoded magnitude spectra with the phase spectra, and by applying inverse STFT with overlap-add. If the latent coefficients are not modified in between encoding and decoding, the decoded magnitude spectra are close to the original ones and the original phase spectra can be directly used for good quality waveform reconstruction. If the latent coefficients are modified so that the decoded magnitude spectra become different from the original one, then the Griffin \& Lim algorithm \cite{Griffin1984} is used to estimate/refine the phase spectra (the original phase spectra are used for initialization) and finally reconstruct the time-domain signal. A few more technical details regarding data pre-processing are given in Section~\ref{sec:pre-proc}. 

%LG: if we modify the latent vector for interpolation, we must talk about phase reconstruction (G&L) here! = present the two cases.

%\begin{figure*}[ht!]
%\begin{minipage}{\textwidth}
%\centering
	%\includegraphics[width=0.7\textwidth]{ISMIR_method_schematic.PNG}
	%\caption{Global diagram of the sound analysis-transformation-synthesis process.}
	%\label{fig:schema}
%\end{minipage}
%\end{figure*}

\subsection{Dimensionality Reduction Techniques}

%As presented in the introduction, both linear (PCA) and non-linear (AE, DAE, VAE) dimensionality reduction techniques were investigated during this work. We present them briefly.

\textbf{Principal Component Analysis}: As a baseline, we investigated the use of PCA to reduce the dimensionality of the input vector $\mathbf{x}$. PCA is the optimal linear orthogonal transformation that provides a new coordinate system (i.e.~the latent space) in which basis vectors
follow modes of greatest variance in the original data \cite{BishopBook}.% TH : j'ai compressé un peu
% is a linear technique commonly used for dimensionality reduction. It maps the original data to a lower-dimensional space extracted such that it maximizes the variance. 
% LG: the variance of what? for me, a PCA is a MMSE dimensionnality reduction technique. In fact it is THE optimal MMSE linear dimensionnality reduction technique = it minimizes MSE(X,Y) with Y = ABX (B = encoding matrix (high-to-low dim linear transform) and A = decoding matrix (low-to-high dim linear transform). I think we should present PCA as what it is = a very classical tool that we do not need to present :-) 
%The eigenvectors of the covariance matrix of the data are computed. The ones corresponding to the highest eigenvalues, and hence to the largest variance of the data, are used as direction-vectors of the new abstraction \cite{BishopBook}.
%One advantage of the PCA is that thanks to the eigenvector decomposition, the dimensions of the extracted latent space are orthogonal.

\begin{figure}[ht!]
\centering
\includegraphics[width=\columnwidth]{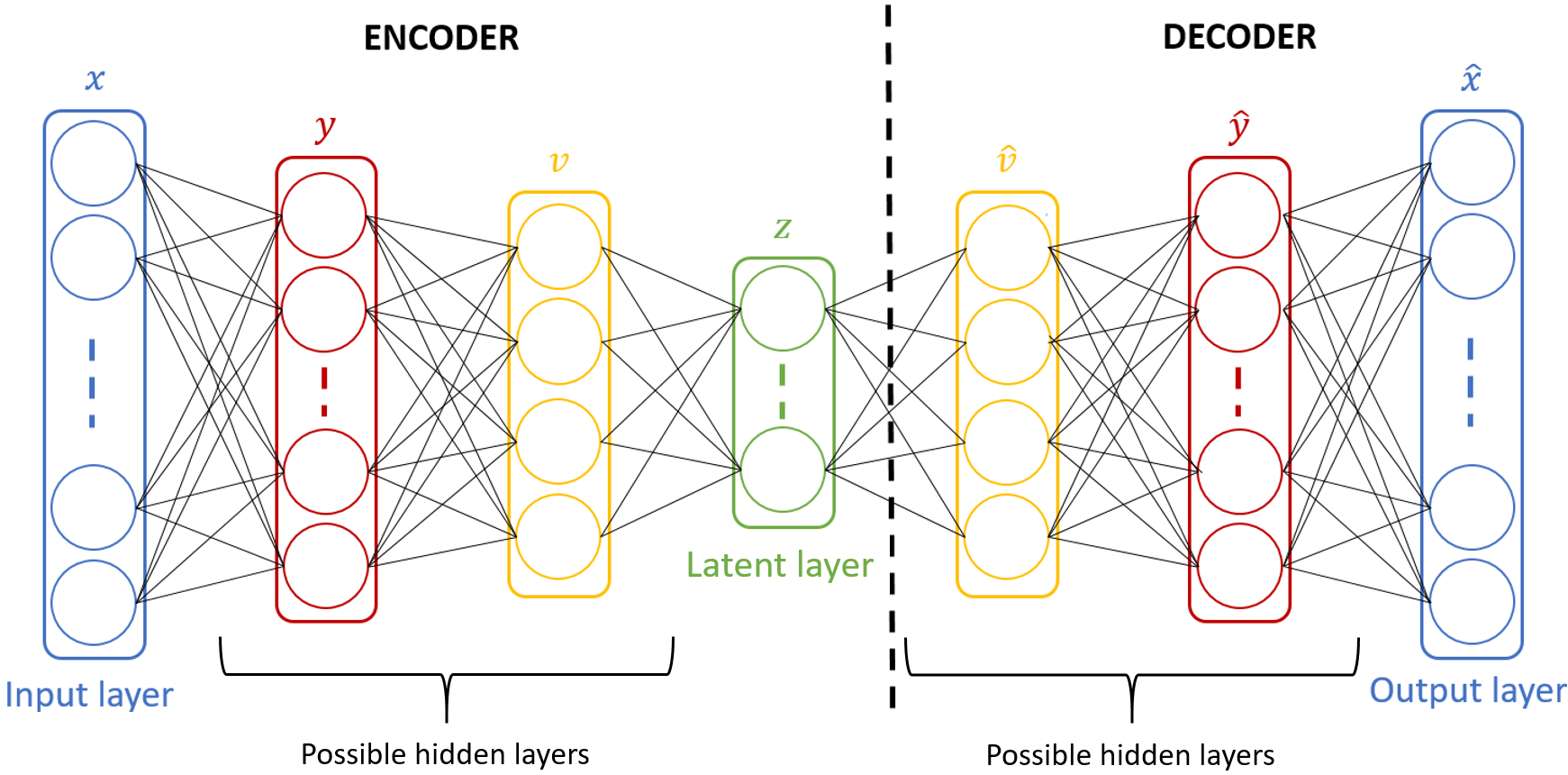}
\caption{General architecture of a (deep) autoencoder.}
\label{fig:autoenc}
\end{figure}

\textbf{Autoencoder}\label{sec:ae}: An AE is a specific kind of ANN traditionally used for dimensionality reduction thanks to its diabolo shape \cite{DLBook}, see Fig.~\ref{fig:autoenc}. It is composed of an encoder and a decoder. The encoder maps a high-dimensional low-level input  vector $\mathbf{x}$ into a low-dimensional higher-level latent vector $\mathbf{z}$, which is assumed to nicely encode properties or attributes of $\mathbf{x}$. Similarly, the decoder reconstructs an estimate $\mathbf{\hat{x}}$ of the input vector $\mathbf{x}$ from the latent vector $\mathbf{z}$. The model is written as: 
\begin{equation*} 
\mathbf{z} = f_{\text{enc}}(\mathbf{W}_{\text{enc}} \mathbf{x} + \mathbf{b}_{\text{enc}}) \quad \textnormal{ and }
\quad \mathbf{\hat{x}} = f_{\text{dec}}(\mathbf{W}_{\text{dec}} \mathbf{z} + \mathbf{b}_{\text{dec}}), 
\end{equation*} 
where $f_{\text{enc}}$ and $f_{\text{dec}}$ are (entry-wise) non-linear activation functions, $\mathbf{W}_{\text{enc}}$ and $\mathbf{W}_{\text{dec}}$ are weight matrices and $\mathbf{b}_{\text{enc}}$ and $\mathbf{b}_{\text{dec}}$ are bias vectors.
For regression tasks (such as the one considered in this study), a linear activation function is generally used for the output layer.

At training time, the weight matrices and the bias vectors are learned by minimizing some cost function over a training dataset. Here we consider the mean squared error (MSE) between the input $\mathbf{x}$ and the output $\mathbf{\hat{x}}$. 
%\begin{equation*} L(\mathbf{\hat{x}}, \mathbf{x}) = \frac{1}{N} \sum\limits_{n=1}^N(\hat{x}[n] - x[n])^2. \end{equation*} 

The model can be extended by adding hidden layers in both the encoder and decoder to create a so-called deep autoencoder (DAE), as illustrated in Fig.~\ref{fig:autoenc}. %This aims at extracting even more abstract latent features. In this case,  end-to-end training of the deep model can be difficult, as it can get stuck in a local minimum. A solution is to proceed to a layer-wise training 
This kind of architecture can be trained globally (end-to-end) or layer-by-layer by considering the DAE as a stack of shallow AEs \cite{Hinton2006, Bengio2007}.

\textbf{LSTM Autoencoder}\label{sec:lstm}: In a general manner, a recurrent neural network (RNN) is an ANN where the output of a given hidden layer does not depend only on the output of the previous layer (as in a feedforward architecture) but also on the internal state of the network. Such internal state can be defined as the output of each hidden neuron when processing the previous input observations.
%layers are composed of recurrent cells that input both the output of the preceding layer at current time and their own output at previous time. 
They are thus well-suited to process time series of data and capture their time dependencies. 
%The motivation for the use of RNNs here is to evaluate if such networks can efficiently capture and reproduce/generate dynamics of (elementary) music sounds.
Such networks are here expected to extract latent representations that encode some aspects of the sound dynamics.
Among different existing RNN architectures, in this study we used the Long Short-Term Memory (LSTM) network \cite{hochreiter1997lstm}, which is known to tackle correctly the so-called vanishing gradient problem in RNNs \cite{bengio1994learning}. 
%, and that are now widely used for the modeling of time sequences. % TH : j'ai également reformulé un peu ce paragraphe
%The LSTM model introduces a cell state in addition to the usual internal state. The different states can be updated thanks to different gates (input, forget and output gates). For more details about LSTM cells and networks, see \cite{hochreiter1997lstm} and subsequent papers by Schmidhuber and colleagues. 
The structure of the model depicted in Fig.~\ref{fig:autoenc} still holds while replacing the classical neuronal cells by LSTM cells, leading to a LSTM-AE. The cost function to optimize remains the same, i.e. the MSE between the input $\mathbf{x}$ and the output $\mathbf{\hat{x}}$. However, the model is much more complex and has more parameters to train \cite{hochreiter1997lstm}. 

\textbf{Variational Autoencoder}\label{sec:vae}: %As mentioned in the introduction, 
A VAE can be seen as a probabilistic AE which delivers a parametric model of the data distribution, such as:
\begin{equation*} 
p_\theta(\mathbf{x},\mathbf{z}) = p_\theta(\mathbf{x}|\mathbf{z})p_\theta(\mathbf{z}),
\end{equation*} 
where $\theta$ denotes the set of distribution parameters. In the present context, the likelihood function $p_\theta(\mathbf{x}|\mathbf{z})$ plays the role of a probabilistic decoder which models how the generation of observed data $\mathbf{x}$ is conditioned on the latent data $\mathbf{z}$. The prior distribution $p_\theta(\mathbf{z})$ is used to structure (or regularize) the latent space.  
Typically a standard Gaussian distribution $p_\theta (\mathbf{z}) = \mathcal{N}(\mathbf{z};\mathbf{0},\mathbf{I})$ is used,
where $\mathbf{I}$ is the identity matrix \cite{Kingma2013}. This encourages the latent coefficients to be mutually orthogonal and lie on a similar range. Such properties may be of potential interest for using the extracted latent coefficients as control parameters of a music sound generator.
%, two interesting properties for a control space for data generation/transformation.
The likelihood $p_\theta(\mathbf{x}|\mathbf{z})$ is defined as a Gaussian density: 
\begin{equation*} 
\label{eq:likelihood}
p_\theta(\mathbf{x}|\mathbf{z}) = \mathcal{N}(\mathbf{x};\boldsymbol{\mu}_\theta(\mathbf{z}), \boldsymbol{\sigma}^2_\theta(\mathbf{z})), 
\end{equation*}
where $\boldsymbol{\mu}_\theta(\mathbf{z})$ and $\boldsymbol{\sigma}^2_\theta(\mathbf{z})$ are the outputs of the decoder network (hence $\theta=\{\mathbf{W}_{\text{dec}},\mathbf{b}_{\text{dec}}\}$).
%\footnote{Note that we can vary the number of layers in the VAE encoder and decoder. In this paper, for clarity, we do not differentiate shallow VAE and deep VAE.} 
Note that $\boldsymbol{\sigma}_\theta^2(\mathbf{z})$ indifferently denotes the covariance matrix of the distribution, which is assumed diagonal, or the vector of its diagonal entries.

The exact posterior distribution $p_\theta(\mathbf{z}|\mathbf{x})$ corresponding to the above model is intractable. It is approximated with a tractable parametric model $q_\phi(\mathbf{z}|\mathbf{x})$ that will play the role of the corresponding probabilistic encoder. This model generally has a form similar to the decoder:  
\begin{equation*} 
q_\phi(\mathbf{z}|\mathbf{x}) = \mathcal{N}(\mathbf{z};\tilde{\boldsymbol{\mu}}_\phi(\mathbf{x}), \tilde{\boldsymbol{\sigma}}^2_\phi(\mathbf{x})), \end{equation*}
where $\tilde{\boldsymbol{\mu}}_\phi(\mathbf{x})$ and $\tilde{\boldsymbol{\sigma}}^2_\phi(\mathbf{x})$ are the outputs of the encoder ANN (the parameter set $\phi$ is composed of $\mathbf{W}_{enc}$ and $\mathbf{b}_{enc}$; $\tilde{\boldsymbol{\sigma}}_\phi^2(\mathbf{x})$ is a diagonal covariance matrix or is the vector of its diagonal entries).

Training of the VAE model, i.e.~estimation of $\theta$ and $\phi$, is done by maximizing the marginal log-likelihood $\textnormal{log}\,p_\theta(\mathbf{x})$ over a large training dataset of vectors $\mathbf{x}$. It can be shown that the marginal log-likelihood can be written as \cite{Kingma2013}:
\begin{equation*} 
\textnormal{log} \ p_\theta(\mathbf{x}) = \textnormal{D}_\textnormal{KL}(q_\phi(\mathbf{z}|\mathbf{x}) | p_\theta(\mathbf{z}|\mathbf{x})) + \mathcal{L}(\phi, \theta, \mathbf{x}),
\end{equation*}  
where $\textnormal{D}_\textnormal{KL} \geq 0 $ denotes the Kullback-Leibler divergence (KLD) and $\mathcal{L}(\phi, \theta, \mathbf{x})$ is the variational lower bound (VLB) given by:
\begin{equation} 
\label{eq:vlb}
\mathcal{L}(\phi, \theta, \mathbf{x}) = \underbrace{- \textnormal{D}_\textnormal{KL}( q_\phi(\mathbf{z}|\mathbf{x}) | p_\theta(\mathbf{z}))}_{\textnormal{regularization}}  +   \underbrace{\mathbb{E}_{q_\phi(\mathbf{z}|\mathbf{x})} [\textnormal{log}\,p_\theta(\mathbf{x}|\mathbf{z})]}_{\textnormal{reconstruction accuracy}}. 
\end{equation} 
In practice, the model is trained by maximizing $\mathcal{L}(\phi, \theta, \mathbf{x})$ over the training dataset with respect to parameters $\phi$ and $\theta$. We can see that the VLB is the sum of two terms. The first term acts as a regularizer encouraging the approximate posterior $q_\phi(\mathbf{z}|\mathbf{x})$ to be close to the prior $p_\theta(\mathbf{z})$. The second term represents the average reconstruction accuracy. Since the expectation w.r.t. $q_\phi(\mathbf{z}|\mathbf{x})$ is difficult to compute analytically, it is approximated using a Monte Carlo estimate and samples drawn from $q_\phi(\mathbf{z}|\mathbf{x})$. For other technical details that are not relevant here, the reader is referred to \cite{Kingma2013}.

As discussed in \cite{blaauw2016modeling} and \cite{Higgins2017}, a weighting factor, denoted $\beta$, can be introduced in (\ref{eq:vlb}) to balance the regularization and reconstruction terms:
\begin{align} 
\label{eq:wvlb}
\mathcal{L}(\phi, \theta, \beta, \mathbf{x}) &= - \beta\,\textnormal{D}_\textnormal{KL}( q_\phi(\mathbf{z}|\mathbf{x}) | p_\theta(\mathbf{z})) \nonumber \\ 
&\qquad \quad + \mathbb{E}_{q_\phi(\mathbf{z}|\mathbf{x})} [\textnormal{log}\,p_\theta(\mathbf{x}|\mathbf{z})], 
\end{align}  
This enables the user to better control the trade-off between output signal quality and compactness/orthogonality of the latent coefficients $\mathbf{z}$. Indeed, if the reconstruction term is too strong relatively to the regularization term, then the distribution of the latent space will be poorly constrained by the prior $p_\theta(\mathbf{z})$, turning the VAE into an AE. Conversely, if it is too weak, then the model may focus too much on constraining the latent coefficients to follow the prior distribution while providing poor signal reconstruction \cite{Higgins2017}. In the present work we used this type of $\beta$-VAE and we present the results obtained with different values of $\beta$. These latter were selected manually after pilot experiments to ensure that the values of the regularization and the reconstruction accuracy terms in (\ref{eq:wvlb}) are in the same range.

%%%%%%%%%%%%%%%%% new section %%%%%%%%%%%%%%%%%%%%%%%%%%%%%%%%%%%%%%%%%%%

\section{Experiments}\label{sec:exp}

\subsection{Dataset}

In this study, we used the NSynth dataset introduced in \cite{NSynth}. This is a large database (more than 30 GB) of 4s long monophonic music sounds sampled at 16 kHz. They represent $1,\!006$ different instruments generating notes with different pitches (from MIDI 21 to 108) and different velocities (5 different levels from 25 to 127). To generate these samples different methods were used: Some acoustic and electronic instruments were recorded and some others were synthesized. The dataset is labeled with: i)~instrument family (e.g., keyboard, guitar, synth\_lead, reed), ii)~source (acoustic, electronic or synthetic), iii)~instrument index within the instrument family, iv)~pitch value, and v)~velocity value. Some other labels qualitatively describe the samples, e.g.~brightness or distortion, but they were not used in our work.

To train our models, we used a subset of $10,\!000$ different sounds randomly chosen from this NSynth database, representing all families of instruments, different pitches and different velocities. We split this dataset into a training set (80\%) and testing set (20\%). During the training phase, 20\% of the training set was kept for validation. 
%All the results presented in this section were obtained using $k$-fold cross-validation with $k=5$.
In order to have a statistically robust evaluation, a $k$-fold cross-validation procedure with $k=5$ was used to train and test all different models (we divided the dataset into $5$ folds, used $4$ of them for training and the remaining one for test, and repeated this procedure $5$ times so that each sound of the initial dataset was used once for testing).

\subsection{Data Pre-Processing}
\label{sec:pre-proc}
For magnitude and phase short-term spectra extraction, we applied a $1,\!024$-point STFT to the input signal using a sliding Hamming window with $50$\% overlap. Frames corresponding to silence segments were removed.
The corresponding $513$-point positive-frequency magnitude spectra were then converted to log-scale and normalized in energy: We fixed the maximum of each log-spectrum input vector to 0\,dB (the energy coefficient was stored to be used for signal reconstruction). Then, the log-spectra were thresholded, i.e.~every log-magnitude below a fixed threshold was set to the threshold value. Finally they were normalized between $-1$ and $1$, which is a usual procedure for ANN inputs. Three threshold values were tested: $-80$~dB, $-90$~dB and $-100$~dB. Corresponding denormalization, log-to-linear conversion and energy equalization were applied after the decoder, before signal reconstruction with transmitted phases and inverse STFT with overlap-add.

\subsection{Autoencoder Implementations}

We tried different types of autoencoders: AE, DAE, LSTM-AE and VAE. For all the models we investigated several values for the encoding dimension, i.e. the size of the bottleneck layer / latent variable vector, from $enc=4$ to $100$ (with a fine-grained sampling for $enc \leq 16$). Different architectures were tested for the DAEs: [513, 128, $enc$, 128, 513], [513, 256, $enc$, 256, 513] and [513, 256, 128, $enc$, 128, 256, 513]. Concerning the LSTM-AE, our implementation used two vanilla forward LSTM layers (one for the encoder and one for the decoder) with non-linear activation functions giving the following architecture: [513, $enc$, 513]. Both LSTM layers were designed for many-to-many sequence learning, meaning that a sequence of inputs, i.e. of spectral magnitude vectors, is encoded into a sequence of latent vectors of same temporal size and then decoded back to a sequence of reconstructed spectral magnitude vectors.
The architecture we used for the VAE was [513, 128, $enc$, 128, 513] and we tested different values of the weight factor $\beta$. For all the neural models, we tested different pairs of activation functions for the hidden layers and output layer, respectively: (tanh, linear), (sigmoid, linear) and (tanh, sigmoid).

AE, DAE, LSTM-AE and VAE models were implemented using the \textit{Keras} toolkit \cite{Keras} (we used the \textit{scikit-learn} \cite{scikit-learn} toolkit for the PCA). Training was performed using the Adam optimizer \cite{Adam} with a learning rate of $10^{-3}$ over $600$ epochs with early stopping criterion (with a patience of $30$ epochs) and a batch size of $512$. The DAEs were trained in two different ways, with and without layer-wise training.
% TH : rajouter la ref de KERAS et de scikit learn. 
% FR: Done !

\subsection{Experimental Results for Analysis-Resynthesis}\label{sec:res}

\begin{figure*}[!ht]
	\begin{multicols}{2}
		\includegraphics[width=0.47\textwidth]{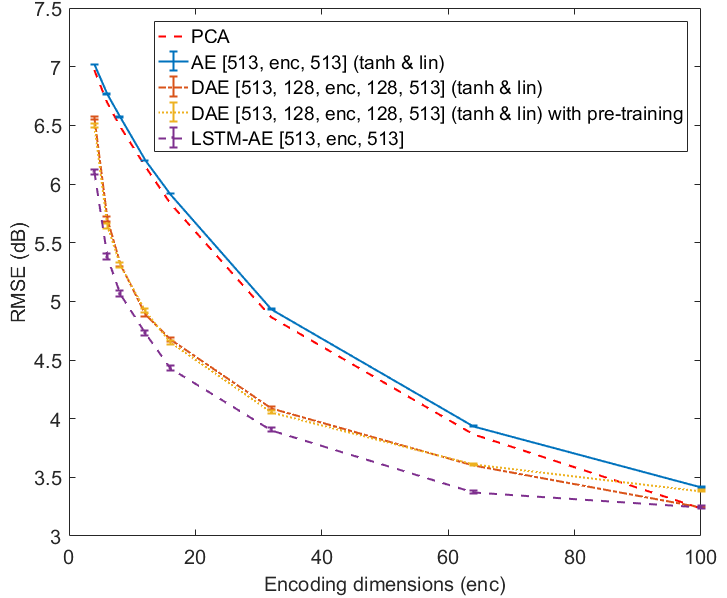}
		\caption{Reconstruction error (RMSE in dB) obtained with PCA, AE, DAE (with and without layer-wise training) and LSTM-AE, as a function of latent space dimension.}
		\label{fig:rmse-AEs}
		\includegraphics[width=0.47\textwidth]{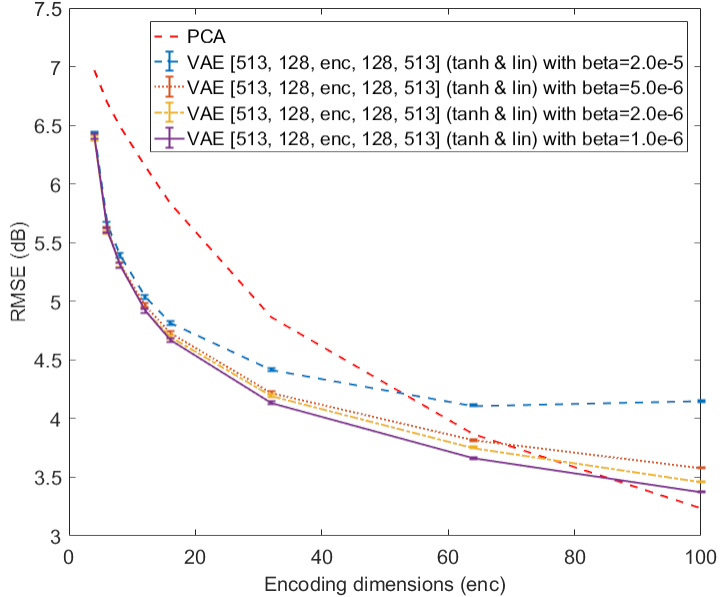}
		\caption{Reconstruction error (RMSE in dB) obtained with VAEs as a function of latent space dimension (RMSE obtained with PCA is also recalled).}
		\label{fig:rmse-VAEs}
    \end{multicols}
\end{figure*}

Fig.~\ref{fig:rmse-AEs} shows the reconstruction error (RMSE in dB) obtained with PCA, AE, DAE and LSTM-AE models on the test set (averaged over the 5 folds of the cross-validation procedure), as a function of the dimension of the latent space. The results obtained with the VAE (using the same protocol, and for different $\beta$ values) are shown in Fig.~\ref{fig:rmse-VAEs}. 
For the sake of clarity, we present here only the results obtained for i) a threshold of $-100$~dB applied on the log-spectra, and ii) a restricted set of the tested AE, DAE and VAE architectures (listed in the legends of the figures). Similar trends were observed for other thresholds and other tested architectures. For each considered dimension of the latent space, a 95\% confidence interval of each reconstruction error was obtained by conducting paired t-test, considering each sound (i.e.~each audio file) of the test set as an independent sample. 

\begin{figure*}[ht!]
	\begin{multicols}{2}
		\includegraphics[width=0.47\textwidth]{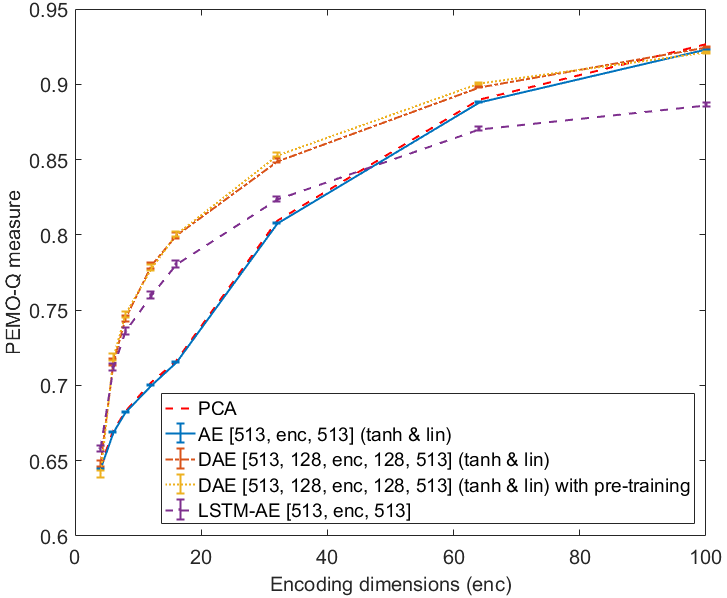} 
		\caption{PEMO-Q measures obtained with PCA, AE, DAEs (with and without layer-wise training) and LSTM-AE, as a function of latent space dimension.}
		\label{fig:pemo-AEs}
		\includegraphics[width=0.47\textwidth]{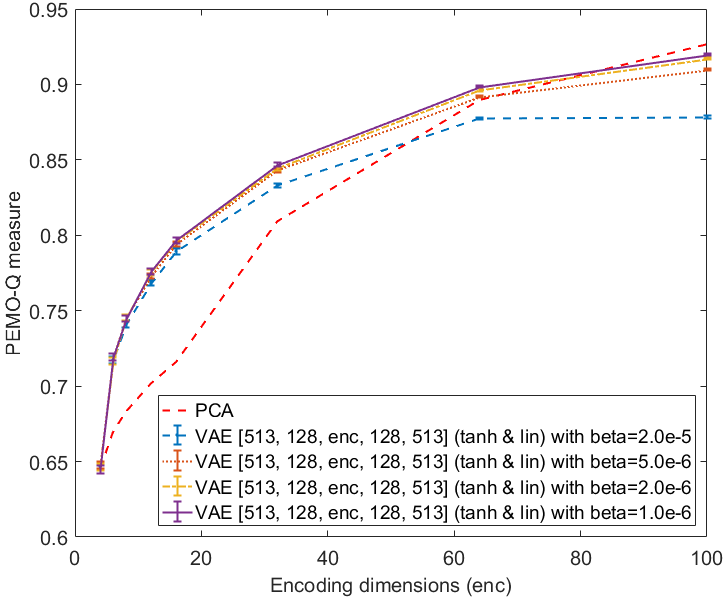}
		\caption{PEMO-Q measures obtained with VAEs as a function of latent space dimension (measures obtained with PCA are also recalled).}
		\label{fig:pemo-VAEs}
    \end{multicols}
\end{figure*}

RMSE provides a global measure of magnitude spectra reconstruction but can be insufficiently correlated to perception depending on which spectral components are correctly or poorly reconstructed. To address this classical issue in audio processing, we also calculated objective measures of perceptual audio quality, namely PEMO-Q scores \cite{huber2006pemo}. The results are reported in Fig.~\ref{fig:pemo-AEs} and Fig.~\ref{fig:pemo-VAEs}.

%\begin{figure}[ht!]
%\centering
%%\includegraphics[width=\columnwidth]{PCAvsAEcomp_thresh100.png} 
%\includegraphics[width=\columnwidth]{RMSE_AE.png} 
%\caption{Reconstruction error (RMSE in dB) obtained with PCA, AE, DAE (with and without layer-wise training) and LSTM-AE, as a function of latent space dimension.}
%\label{fig:rmse-AEs}
%\end{figure}
%
%\begin{figure}[ht!]
%\centering
%%\includegraphics[width=0.85\columnwidth]{PCAvsVAEcomp_thresh100.png} 
%\includegraphics[width=\columnwidth]{RMSE_VAE.png} 
%\caption{Reconstruction error (RMSE in dB) obtained with VAEs as a function of latent space dimension (RMSE obtained with PCA is also recalled).}
%\label{fig:rmse-VAEs}
%\end{figure}
%
%\begin{figure}[ht!]
%\centering
%\includegraphics[width=\columnwidth]{PemoQAE.png} 
%\caption{PEMO-Q measures obtained with PCA, AE, DAEs (with and without layer-wise training) and LSTM-AE, as a function of latent space dimension.}
%\label{fig:pemo-AEs}
%\end{figure}
%
%\begin{figure}[ht!]
%\centering
%\includegraphics[width=\columnwidth]{PemoQVAE.png} 
%\caption{PEMO-Q measures obtained with VAEs as a function of latent space dimension (measures obtained with PCA are also recalled).}
%\label{fig:pemo-VAEs}
%\end{figure}

As expected, the RMSE decreases with the dimension of the latent space for all methods. Interestingly, PCA systematically outperforms (or at worst equals) shallow AE. This somehow contradicts recent studies on image compression for which a better reconstruction is obtained with AE compared to PCA \cite{Hinton2006}.
To confirm this unexpected result, we replicated our PCA vs.~AE experiment on the MNIST image dataset \cite{MNIST}, using the same AE implementation and a standard image preprocessing (i.e.~vectorization of each $28 \times 28$ pixels gray-scale image into a $784$-dimensional feature vector). In accordance with the literature, the best performance was systematically obtained with AE (for any considered dimension of the latent space). This difference of AE's behavior when considering audio and image data was unexpected and, to our knowledge, it has never been reported in the literature.

Then, contrary to (shallow) AE, DAEs systematically outperform PCA (and thus AE), with up to almost 20\% improvement (for $enc = 12$ and $enc=16$). Our experiments did not reveal notable benefit of layer-by-layer DAE training over end-to-end training. Importantly, for small dimensions of the latent space (e.g.~smaller than 16), RMSE obtained with DAE decreases much faster than with PCA and AE. This is even more the case for LSTM-AE which shows an improvement of the reconstruction error of more than 23\% over PCA (for $enc = 12$ and $enc=16$). These results confirm the benefits of using a more complex architecture than shallow AE, here deep or recurrent, to efficiently extract high-level abstractions and compress the audio space. This is of great interest for sound synthesis for which the latent space has to be kept as low-dimensional as possible (while maintaining a good reconstruction accuracy) in order to be ``controlled'' by a musician.  
% TH: c'est un peu bateau mais bon ... 

Fig.~\ref{fig:rmse-VAEs} shows that the overall performance of VAEs is in between the performance of DAEs (even equals DAEs for lower encoding dimensions, say smaller than 12) and the performances of PCA and AE. Let us recall that minimizing the reconstruction accuracy is not the only goal of VAE which also aims at constraining the distribution of the latent space. As shown in Fig.~\ref{fig:rmse-VAEs}, the parameter $\beta$, which balances regularization and reconstruction accuracy in (\ref{eq:wvlb}), plays a major role. As expected, high $\beta$ values foster regularization at the expense of reconstruction accuracy. However, with $\beta \leqslant 2.10^{-6}$ the VAE clearly outperforms PCA, e.g. up to $20\%$ for $enc = 12$. 

It can be noticed that when the encoding dimension is high ($enc = 100$), PCA seems to outperform all the other models. Hence, in that case, the simpler (linear model) seems to be the best (we can conjecture that achieving the same level of performance with autoencoders would require more training data, since the number of free parameters of these model increases drastically). %However, this is out of the scope of possible applications to sound synthesis, since a $100$-dimensional space is too high-dimensional to be easily controlled by a musician.
However, using such high-dimensional latent space as control parameters of a music sound generator is impractical.

Similar conclusions can be drawn from Fig.~\ref{fig:pemo-AEs} and Fig.~\ref{fig:pemo-VAEs} in terms of audio quality. Indeed, in a general manner, the PEMO-Q scores are well correlated with RMSE measures in our experiments. PEMO-Q measures for PCA and AE are very close, but PCA still slightly outperforms the shallow AE. The DAEs and the VAEs both outperform the PCA (up to about $11\%$ for $enc =12$ and $enc=16$) with the audio quality provided by the DAEs being a little better than for the VAEs. 
%Surprisingly, for the LSTM-AE only, the PEMO-Q measure does not correlate with RMSE as the curve lies slightly below the DAEs and the VAEs curves whereas it lies slightly above for RMSE. This results were also unexpected from the qualitative listening experience. Further investigations will be done to understand this phenomenon.
%commentaire sur le LSTM !
Surprisingly, and contrary to RMSE scores, the LSTM-AE led to a (slightly) lower PEMO-Q scores, for all considered latent dimensions. Further investigations will be done to assess the relevance of such differences at the perceptual level. 

\subsection{Decorrelation of the Latent Dimensions}\label{sec:decorr}

\begin{figure}[ht!]
	\centering
	\includegraphics[width=\linewidth]{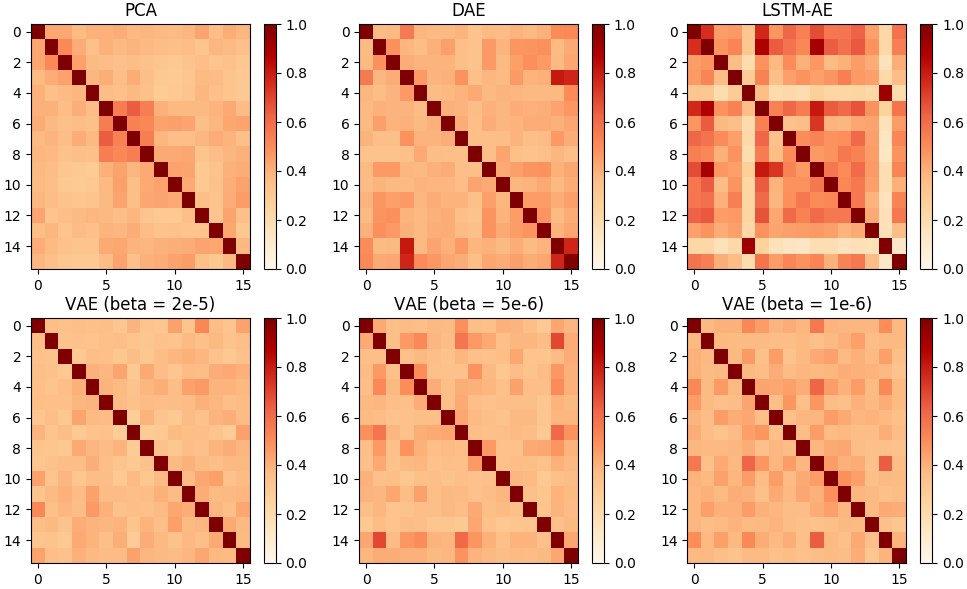}
	\caption{Correlation matrices of the latent dimensions (average absolute correlation coefficients) for PCA, DAE, LSTM-AE and VAEs.}
	\label{fig:corrcoef}
\end{figure}

Now we report further analyses aiming at investigating how the extracted latent dimensions may be used as \textit{control} parameters by the musician. In the present sound synthesis framework, such control parameters are expected to respect (at least) the following two constraints i) to be as decorrelated as possible in order to limit the redundancy in the spectrum encoding, ii) to have a clear and easy-to-understand perceptual meaning. In the present study, we focus on the first constraint by comparing PCA, DAEs, LSTM-AE and VAEs in terms of correlation of the latent dimensions. More specifically, the absolute values of the correlation coefficient matrices of the latent vector $\mathbf{z}$ were computed on each sound from the test dataset and Fig.~\ref{fig:corrcoef} reports the mean values averaged over all the sounds of the test dataset. For the sake of clarity, we present here these results only for a latent space of dimension $16$ for one model of DAE ([513, 128, 16, 128, 513] (tanh \& lin) with end-to-end training) and for VAEs with the same architecture ([513, 128, 16, 128, 513] (tanh \& lin)) and different values of $\beta$ (from $1.10^{-6}$ to $2.10^{-5}$). 

As could be expected from the complexity of its structure, we can see that the LSTM-AE extracts a latent space where the dimensions are significantly correlated with each other. 
Such additional correlations may come from the sound dynamics which provide redundancy in the prediction.
We can also see that PCA and VAEs present similar behaviors with much less correlation of the latent dimensions, which is an implicit property of these models. Interestingly, and in accordance with (\ref{eq:wvlb}), we can notice that the higher the $\beta$, the more regularized the VAE and hence the more decorrelated the latent dimensions. Importantly, Fig.~\ref{fig:corrcoef} clearly shows that for a well-chosen $\beta$ value, the VAE can both extract latent dimensions that are much less correlated than for corresponding DAEs, which makes it a better candidate for extracting good control parameters, while allowing fair to good reconstruction accuracy (see Fig.~\ref{fig:rmse-VAEs}). The $\beta$ value has thus to be chosen wisely in order to find the optimal trade-off between decorrelation of the latent dimensions and reconstruction accuracy.

%\begin{figure}[ht!]
	%\centering
	%\includegraphics[width=\linewidth]{Corrcoef_comp_AE.png}
	%\caption{Comparison of absolute correlation coefficients matrices of the latent dimensions of VAEs for different beta values.}
	%\label{fig:corrcoef_AE}
%\end{figure}

\subsection{Examples of Sound Interpolation}\label{sec:interp}

As a first step towards the practical use of the extracted latent space for navigating through the sound space and creating new sounds, we illustrate how it can be used to interpolate between sounds, in the spirit of what was done for instrument hybridization in \cite{NSynth}. We selected a series of pairs of sounds from the NSynth dataset with the two sounds in a pair having different characteristics. For each pair, we proceeded to separate encoding, entry-wise linear interpolation of the two resulting latent vectors, decoding, and finally individual signal reconstruction with inverse STFT and the Griffin and Lim algorithm to reconstruct the phase spectrogram \cite{Griffin1984}. We experimented different degrees of interpolation between the two sounds:  $\hat{\mathbf{z}}=\ \alpha \ \mathbf{z}_1 + (1-\alpha) \ \mathbf{z}_2$, with $\mathbf{z}_i$ the latent vector of sound $i$, $\hat{\mathbf{z}}$ the new interpolated latent vector, and $\alpha \in [0, 0.25, 0.5, 0.75 , 1]$ (this interpolation is processed independently on each pair of vectors of the time sequence).
%$25\%$ of the first sound and $75\%$ of the second, $50\%$-$50\%$ and $75\%$-$25\%$. 
The same process was applied using the different AE models we introduced earlier. 

\begin{figure*}[ht!]
\centering
\begin{multicols}{2}
    \subfloat[Original samples - Left : bass\_electronic\_010-055-100, Right : brass\_acoustic\_050-055-100]{\includegraphics[width=0.45\textwidth]{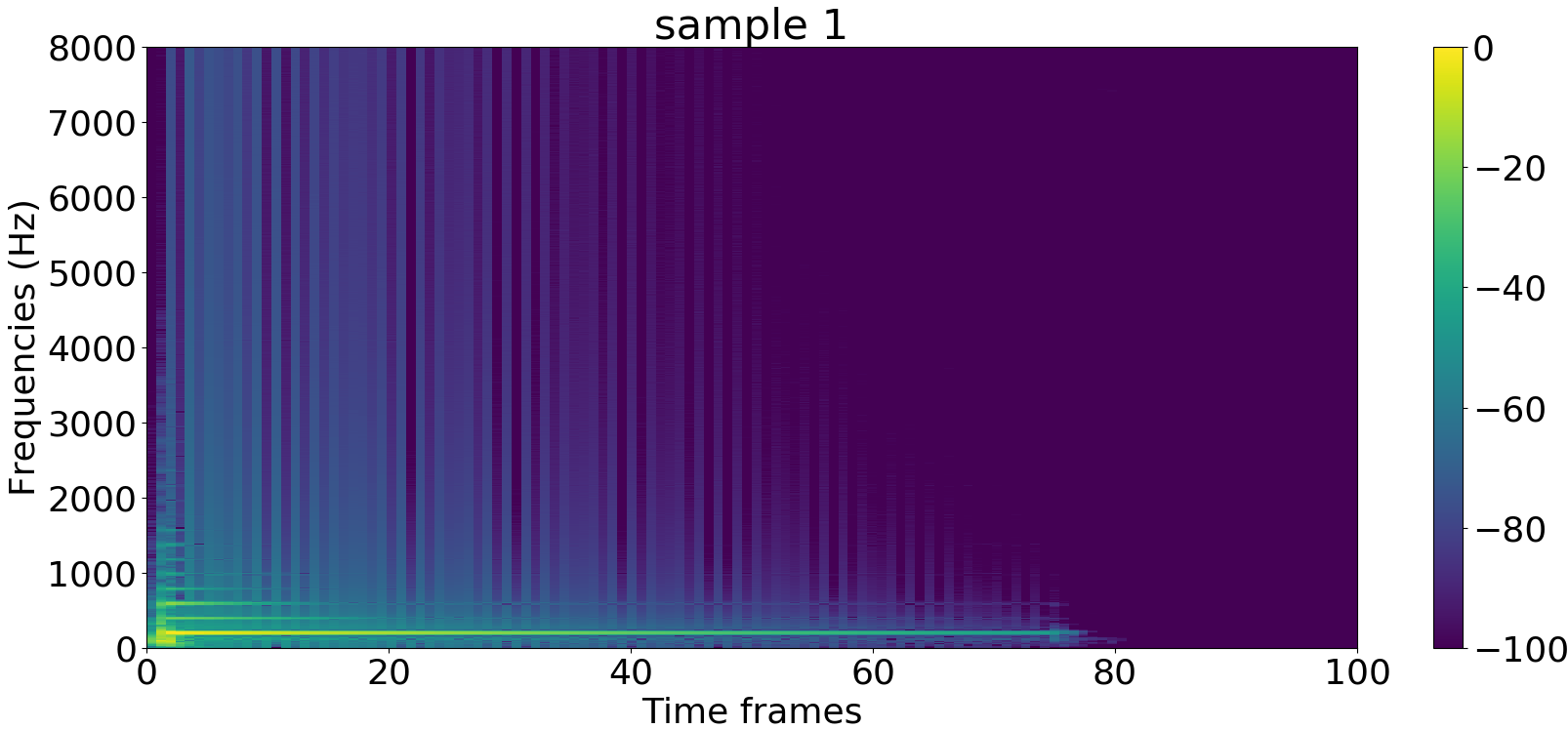}
     \par \hspace{0.08\textwidth} \includegraphics[width=0.45\textwidth]{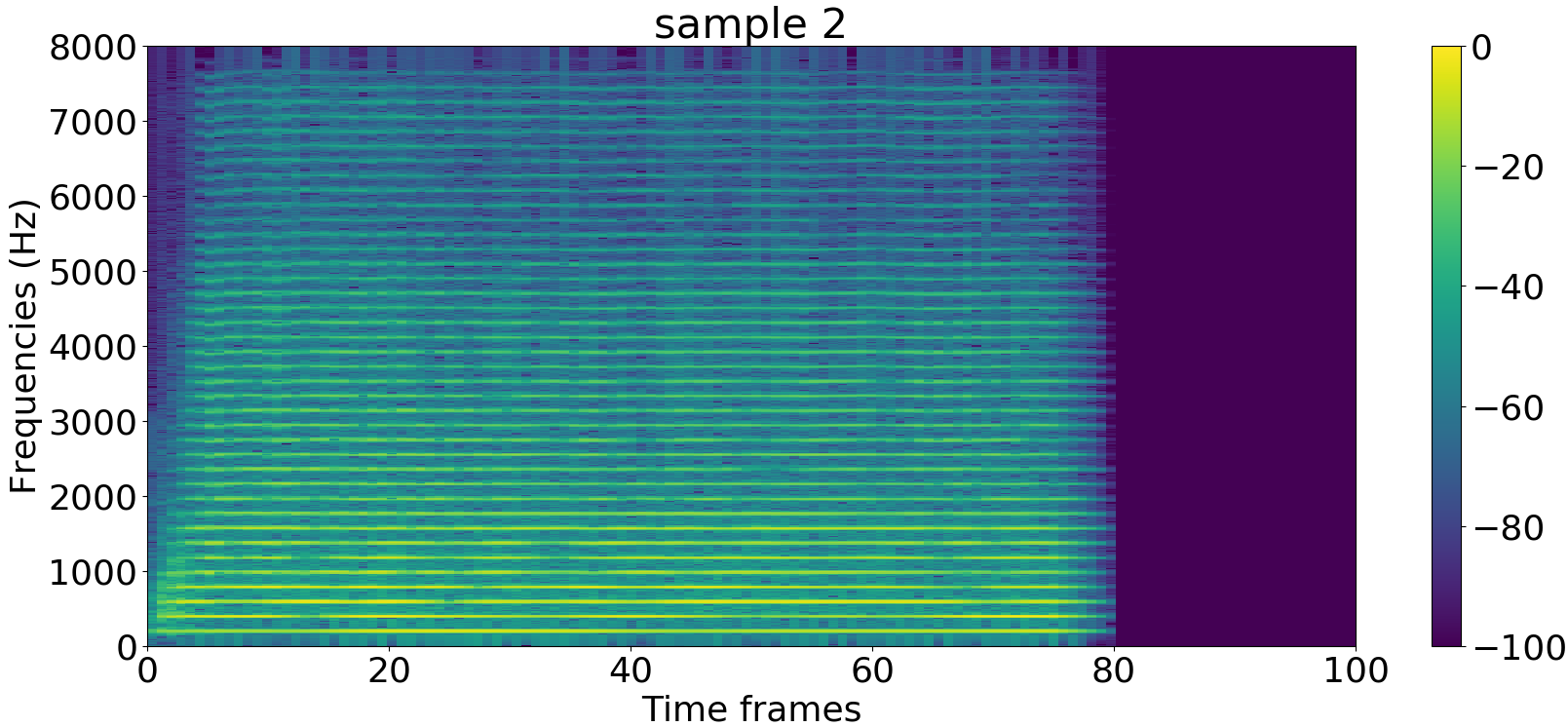} \par }
\end{multicols}
\vspace{-0.8cm}
\begin{multicols}{2}
   \subfloat[PCA]{\includegraphics[width=0.45\textwidth]{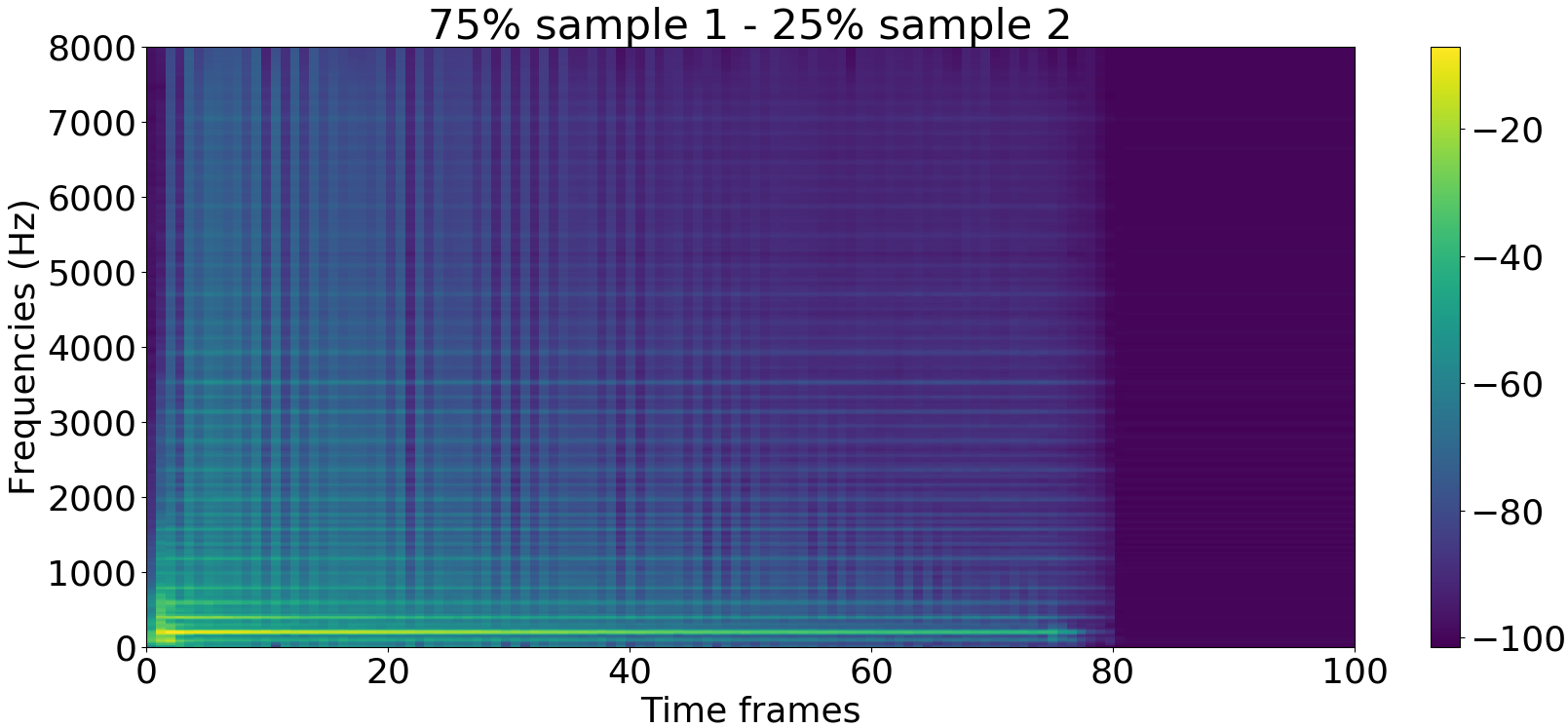}
	 \par \hspace{0.08\textwidth} \includegraphics[width=0.45\textwidth]{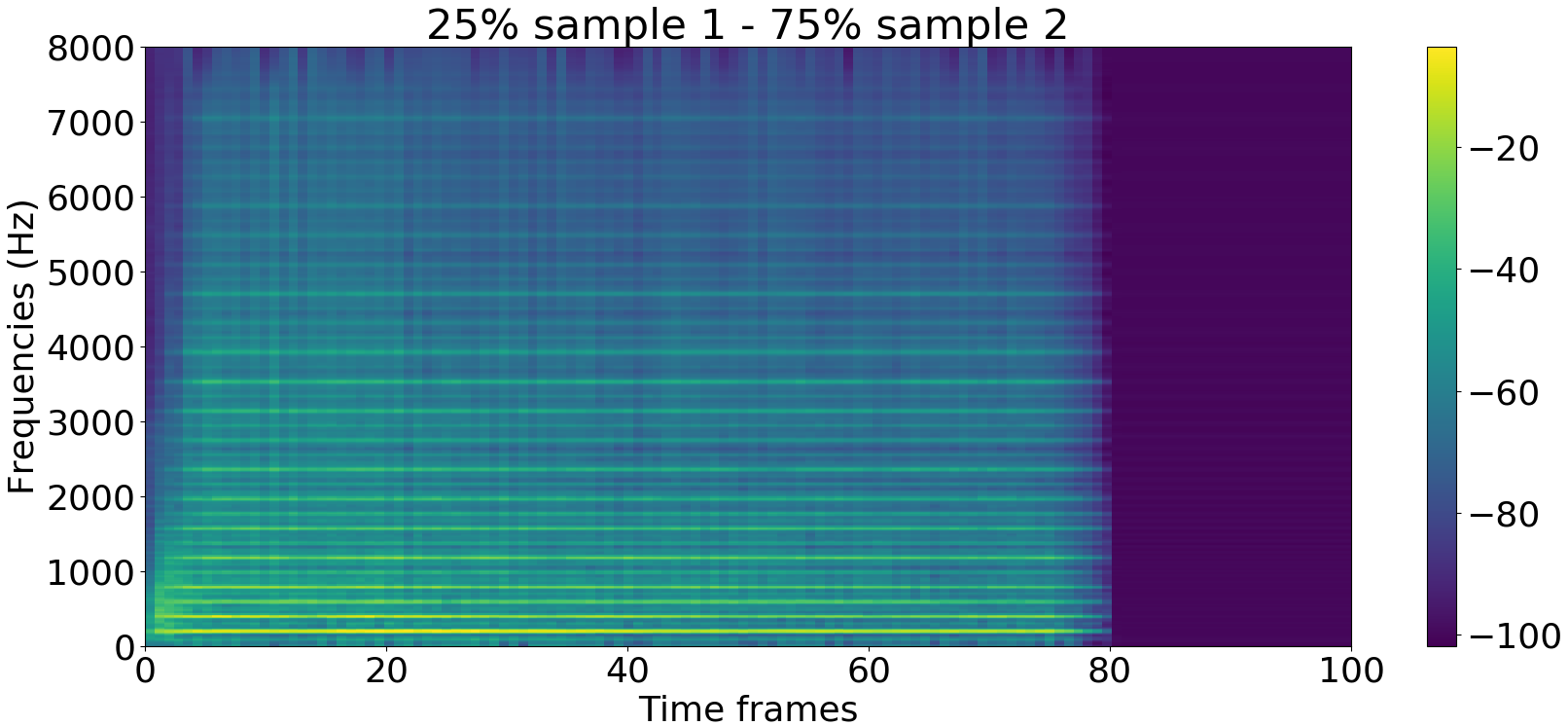} \par } 
    \end{multicols}
    \vspace{-0.8cm}
\begin{multicols}{2}
   \subfloat[LSTM-AE]{\includegraphics[width=0.45\textwidth]{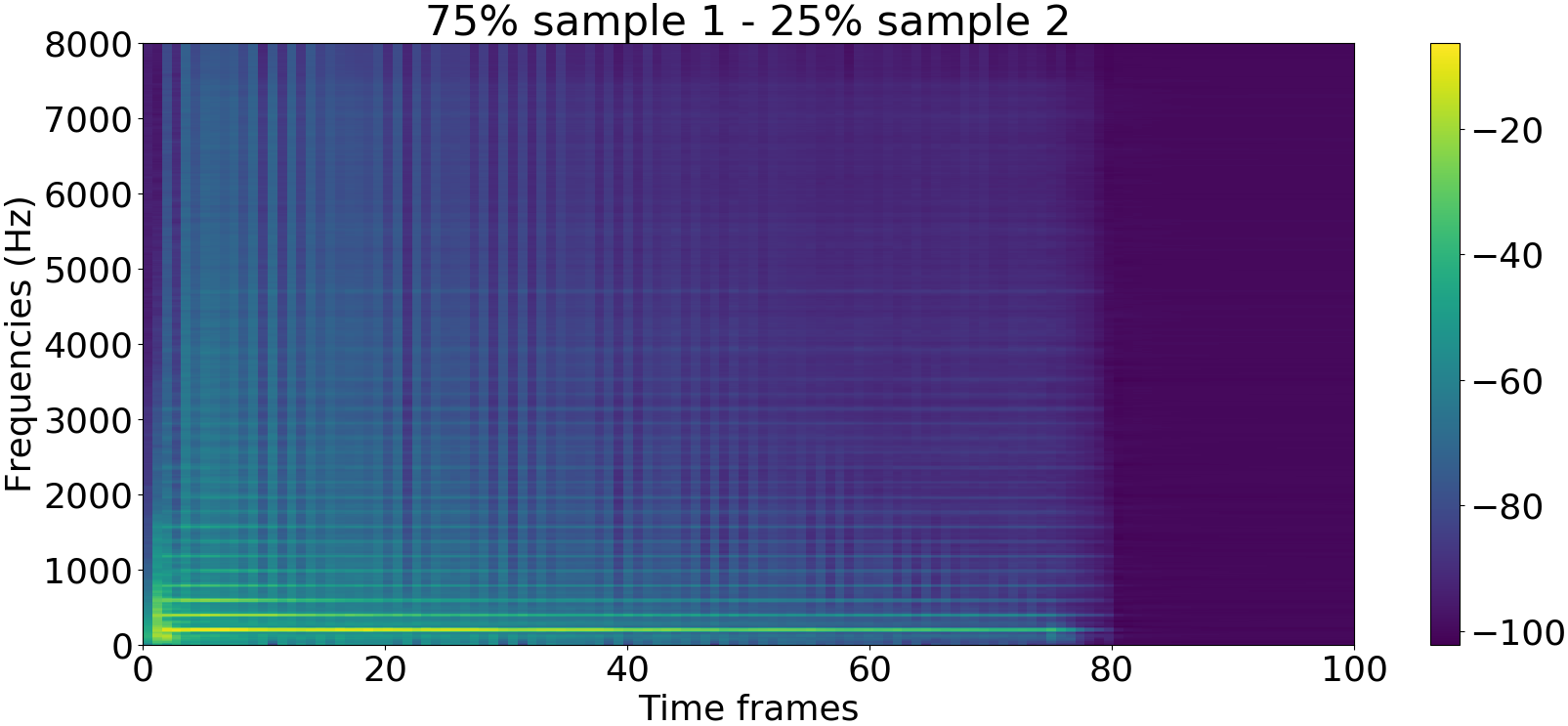}
	 \par \hspace{0.08\textwidth} \includegraphics[width=0.45\textwidth]{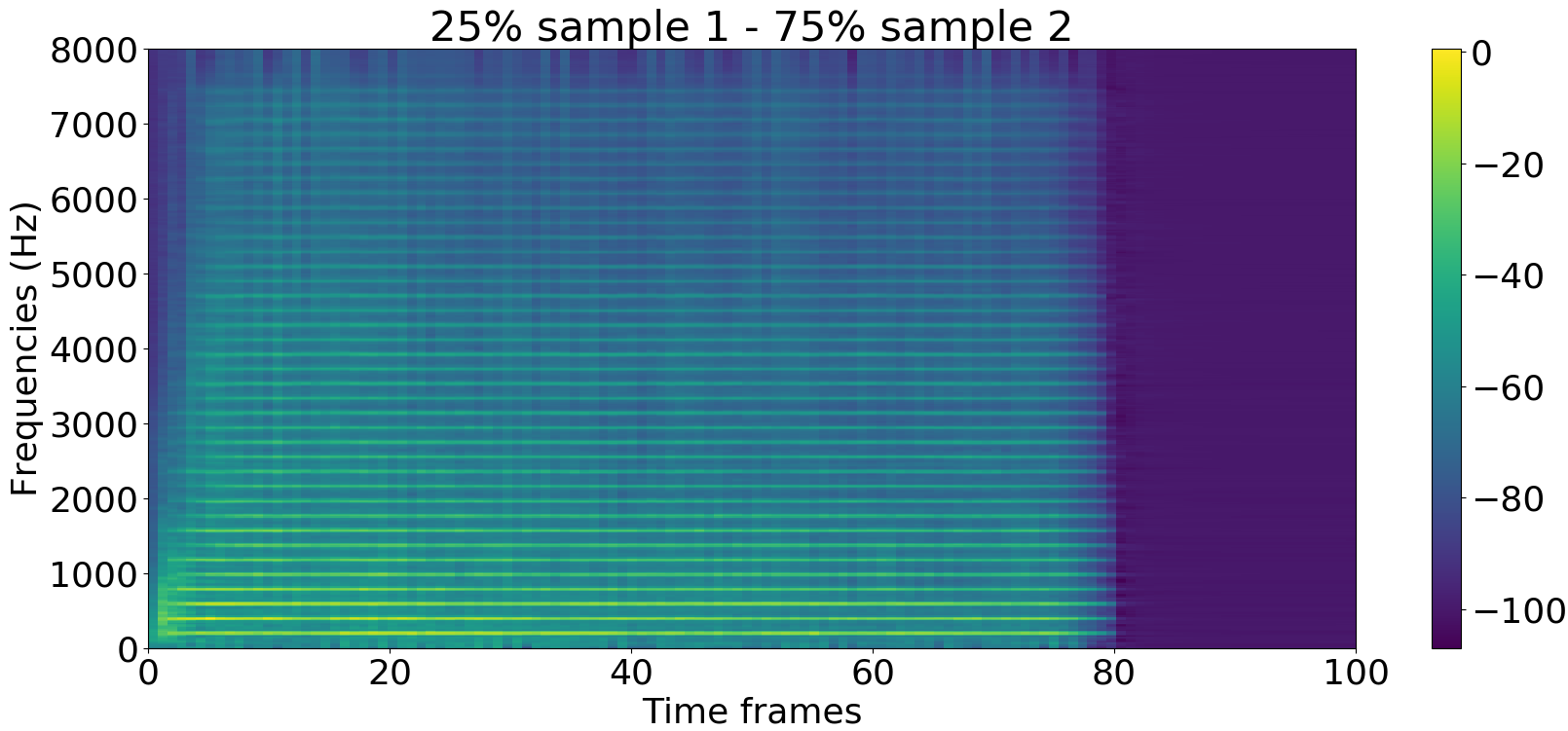} \par } 
    \end{multicols}
    \vspace{-0.8cm}
\begin{multicols}{2}
    \subfloat[VAE]{\includegraphics[width=0.45\textwidth]{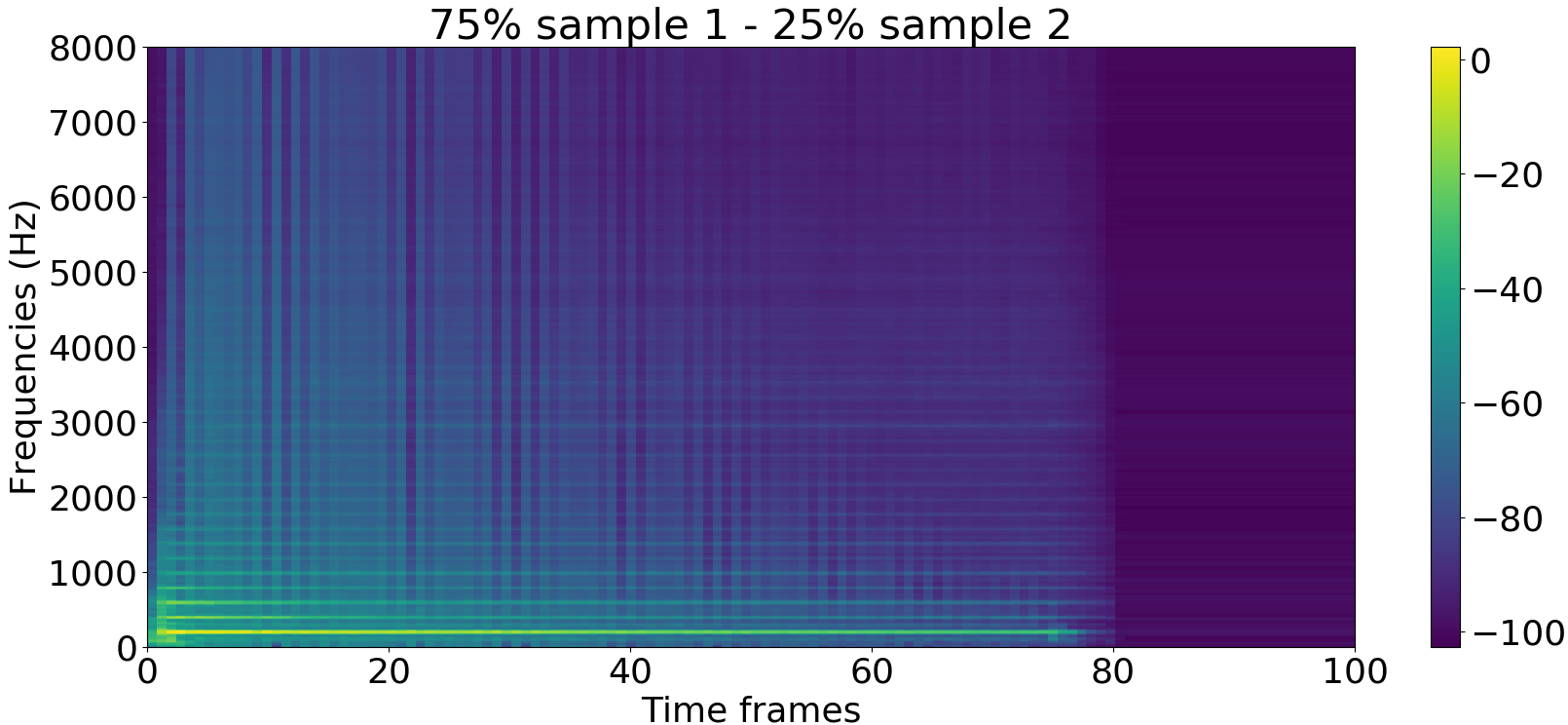}
     \par \hspace{0.08\textwidth} \includegraphics[width=0.45\textwidth]{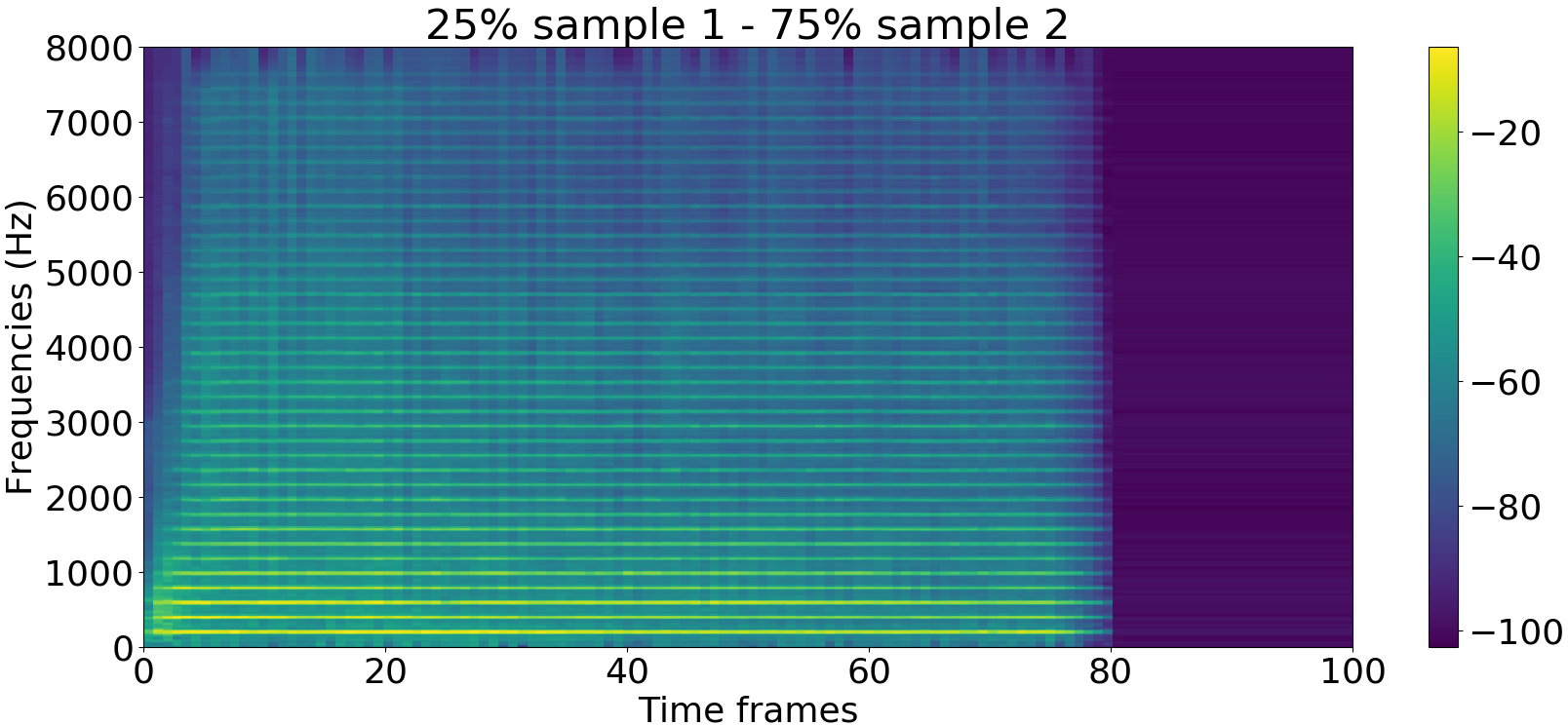} \par }
\end{multicols}
\vspace{-0.5cm}
\caption{Examples of decoded magnitude spectrograms after sound interpolation of 2 samples (top) in the latent space using respectively PCA (2nd row), LSTM-AE (3rd row) and VAE (bottom). A more detailed version of the figure can be found at \url{https://goo.gl/Tvvb9e}.}
\label{fig:interp}
\end{figure*}

Fig.~\ref{fig:interp} displays one example of results obtained with PCA, with the LSTM-AE and with the VAE (with $\beta = 1.10^{-6}$), with an encoding dimension of $32$. Qualitatively, we note that interpolations in the latent space lead to a smooth transition between source and target sound. By increasing sequentially the degree of interpolation, we can clearly go from one sound to another in a consistent manner, and create interesting hybrid sounds. The results obtained using PCA interpolation are (again qualitatively) below the quality of the other models. The example spectrogram obtained with interpolated PCA coefficients is blurrier around the harmonics and some audible artifacts appear. On the opposite, the LSTM-AE seems to outperform the other models by better preserving the note attacks (see comparison with VAE in Fig.~\ref{fig:interp}).
More interpolation examples along with corresponding audio samples can be found at \url{https://goo.gl/Tvvb9e}. %\url{http://www.gipsa-lab.fr/~fanny.roche/SMC_2019.html}.

\section{Conclusions and Perspectives}\label{sec:conclusion}
In this study, we investigated dimensionality reduction based on autoencoders to extract latent dimensions from a large music sound dataset. Our goal is to provide a musician with a new way to generate sound textures by exploring a low-dimensional space. From the experiments conducted on a subset of the publicly available database NSynth, we can draw the following conclusions: i) Contrary to the literature on image processing, shallow autoencoders (AEs) do not here outperform principal component analysis (in terms of reconstruction accuracy); ii) The best performance in terms of signal reconstruction is always obtained with deep or recurrent autoencoders (DAEs or LSTM-AE); iii) Variational autoencoders (VAEs) lead to a fair-to-good reconstruction accuracy while constraining the statistical properties of the latent space, ensuring some amount of decorrelation across latent coefficients and limiting their range. These latter properties make the VAEs good candidates for our targeted sound synthesis application. 

In line with the last conclusion, future works will mainly focus on VAEs. 
First, we will investigate recurrent architecture for VAE such as the one proposed in \cite{Chung2015}. Such approach may lead to latent dimensions encoding separately the sound texture and its dynamics, which may be of potential interest for the musician. % in order to take into account both decorrelation constraints on the extracted latent space and the temporal characteristics of a musical sound. Hopefully, one would like to extract dimensions that more or less independently encode the instrument timbre (perceptually meaningful characteristics of the spectral shape) and the sound dynamics. 
%Finally, experiments will be conducted with Generative Adversarial Networks \cite{Goodfellow2014} which is another popular technique for data-driven deep-learning-based data generation.

Then, we will address the crucial question of the perceptual meaning/relevance of the %resulting 
latent dimensions. Indeed using a non-informative prior distribution of $\mathbf{z}$ such as a standard normal distribution does not ensure that each dimension of $\mathbf{z}$ represents an interesting perceptual dimension of the sound space, although this is a desirable objective. 
%As briefly stated in the introduction, the study \cite{Esling2018} opened research in that direction. The authors introduced in the variational lower bound (\ref{eq:wvlb}) an additional regularization term encouraging the latent space to get close to a space of instrument timbres, as obtained from previous Multi-Dimensional Scaling (MDS) analysis. In this line, we are currently leading a perceptual study to characterize the main perceptual dimensions of synthetic sounds in order to get meaningful and relevant controls for the synthesis by linking these dimensions to the extracted latent space.
In \cite{Esling2018}, the authors recently proposed a first solution to this issue in the context of a restricted set of acoustic instruments. They introduced in the variational lower bound (\ref{eq:wvlb}) of the VAE loss an additional regularization term encouraging the latent space to respect the structure of the instrument timbre. In the same spirit, our future works will investigate different strategies to model the complex relationships between sound textures and their perception, and introduce these models at the VAE latent space level.

% limited range of z à rajouter dans le texte

\section{Acknowledgment}
The authors would like to thank Simon Leglaive for our fruitful discussions. This work was supported by ANRT in the framework of the PhD program CIFRE 2016/0942. 

%%%%%%%%%%%%%%%%%%%%%%%%%%%%%%%%%%%%%%%%%%%%%%%%%%%%%%%%%%%%%%%%%%%%%%%%%%%%%
%bibliography here
\bibliography{rochefa_etal_ismir_2018}

\end{document}